\documentclass[a4paper]{PoS}

\pdfoutput=1
\usepackage{amsmath}
\usepackage[alsoload=hep]{siunitx}
\usepackage{xspace}
\usepackage{float}
\usepackage{subfig}
\usepackage{todonotes}
\usepackage{epstopdf}

\DeclareSIUnit\eVperc{\eV\per\clight}
\newcommand{\nark}{\(\text{NA62}\text{-}{\text{R}_\text{K}}\)\xspace}
\newcommand{\pizerod}{\pi^0_D}
\newcommand{\npcl}{\SI{90}{\percent} CL\xspace}

\title{Recent results from NA48/2 and NA62 experiments at CERN}

\ShortTitle{Recent results from NA48/2 and NA62 experiments at CERN}

\author{\speaker{Nicolas Lurkin}\thanks{Supported by ERC Starting Grant 336581}
\ on behalf of the NA48/2 and NA62 collaboration
\thanks{F.~Ambrosino, A.~Antonelli, G.~Anzivino, R.~Arcidiacono, 
W.~Baldini, S.~Balev, J.R.~Batley, M.~Behler, S.~Bifani, C.~Biino
% ...
}\\
        School of Physics and Astronomy, University of Birmingham\\
        E-mail: \email{nicolas.lurkin@cern.ch}}

\abstract{The NA48/2 and \nark experiments at the CERN SPS collected a large sample of charged kaon decays in
flight. \nark was running in 2007--08 with a highly efficient minimum bias trigger for decays into electrons. A
preliminary measurement of the electromagnetic transition form factor slope of the \(\pi^0\) from \num{1.05e6} fully
reconstructed \(\pi^0\) Dalitz decays is presented. The obtained value \(a = (3.70 \pm 0.53_\text{stat} \pm
0.36_\text{syst})\times 10^{-2}\) represents a \(5.8\sigma\) observation of a non-zero slope in the time-like
region of momentum transfer. 
An upper limit on the rate of a lepton number violating decay \(K^\pm\to\pi^\mp\mu^\pm\mu^\pm\) is
reported from \(\sim\num{1.6e11} K^\pm\) decays at NA48/2 in 2003--04: \(\mathcal{B} < \num{8.6e-11}\) at \npcl.
Searches for heavy sterile neutrino \(N_4\) and inflaton \(\chi\) resonances in \(K^\pm\to\pi\mu\mu\) decays are
reported. No signal is observed and upper limits on the products \(\mathcal{B}(K^\pm\to\mu^\pm
N_4)\mathcal{B}(N_4\to\pi^\mp\mu^\pm)\) and \(\mathcal{B}(K^\pm\to\pi^\pm \chi)\mathcal{B}(\chi\to\mu^+\mu^-)\) are set
in the range \SIrange[range-units=single]{e-10}{e-9} for resonance lifetimes up to \SI{100}{\pico\second}.
The result of a search for dark photon with the same sample of decays is also reported. In the absence of observed
signal, the limits on the mixing parameter \(\varepsilon^2\) in the range \SIrange[range-units=single]{9}{70}{\mega
eV\per\clight\squared} are improved.}

\FullConference{XIII International Conference on Heavy Quarks and Leptons\\
		22-27 May, 2016\\
		Blacksburg, Virginia, USA}

\begin{document}

\section{The NA48/2 and \nark setup}
\label{sec:intro}
The NA48/2 experiment operating in 2003--04, and its successor \nark phase in 2007--08, collected
charged kaon decays in flight at the CERN SPS. The aim of NA48/2 was the measurement of direct CP violation
\cite{Batley2007} with the world's largest sample of charged kaons decays, while \nark aimed at measuring the lepton
universality in the kaon decay \cite{Lazzeroni2013} with a minimum bias data sample. In both cases
the amount of statistics recorded allowed us to study other processes. Lepton number violating decays
and resonance searches in \(K^\pm\to\pi\mu\mu\) and dark photon searches with NA48/2 are reported, as well as the
\(\pi^0\) electromagnetic transition form factor with \nark.

The beam line described in detail in \cite{Batley2007} was designed to provide simultaneous \(K^+\) and
\(K^-\) beams. They were extracted from the \SI{400}{\giga\eVperc} SPS proton beam impinging on a \SI{40}{\cm} long
beryllium target. The final beam momentum of \SI{60(3)}{\giga\eVperc} (NA48/2) and \SI{74.0(14)}{\giga\eVperc}
(\nark) was selected using a system of dipole magnets and a momentum-defining slit incorporated into a beam
dump. This \SI{3.2}{\m} thick copper/iron block provided the possibility to block either of the \(K^+\) or \(K^-\) beams. The
beams were focused and collimated before entering the \SI{114}{\m} long cylindrical vacuum tank containing the fiducial
decay volume. The beam contained mainly pions but included approximately \SI{6}{\percent} of kaons.
The simultaneous \(K^+/K^-\) beams provided about \num{6.2e7} particles per spill of \(\SI{4.8}{\second}\) for
NA48/2, while \nark was also using a single mode beam with an intensity \num{\sim 10} times lower, alternating periods
with \(K^+\), periods with \(K^-\), and periods with both.

The principal subdetectors in use were the same in both cases and a detailed description is given in \cite{Fanti2007}.
The momenta of the charge particles were measured by a spectrometer housed in a tank separated from the decay
volume and filled with helium. It was composed of four drift chambers (DCH) and a dipole magnet between the second and
third ones. The magnet provided a horizontal transverse momentum kick of \SI{120}{\mega\eVperc} in NA48/2 and
\SI{265}{\mega\eVperc} in \nark. A hodoscope (HOD) composed of two planes of plastic scintillator was placed after the
spectrometer to provide precise timing of the charged particles and generate fast trigger signals for the low-level trigger. A
\SI{127}{\cm} thick quasi-homogeneous electromagnetic calorimeter filled with liquid krypton (LKr) was located
downstream and used both as a photon detector and for particle identification. The volume is divided into \num{13248}
cells of \(\sim 2\times 2\SI{}{\;\cm\squared}\) cross section without longitudinal segmentation. The energy resolution
is \(\sigma_E/E = \left(3.2/\sqrt{E} \oplus 9/E \oplus 0.42\right) \percent\) and the position resolution is 
\(\sigma_x~=~\sigma_y~=~\left(4.2/\sqrt{E} \oplus 0.6\right)\si{\mm}\) where the particle energy \(E\) is given 
in \si{\giga\eV}. A muon veto system (MUV) was installed behind the LKr and consisted of three planes of scintillator
orthogonal to the beam axis, each one preceded by a \SI{80}{\cm} thick iron wall. They were made of \SI{2.7}{\m} long
and \SI{2}{\cm} thick strips alternatively arranged horizontally and vertically. The width of the strips
was \SI{25}{\cm} in the first two planes and \SI{45}{\cm} in the last one. The central strips were divided in two
halves to accommodate the beam pipe in a \SI[product-units=power]{22x22}{\cm} central hole. The detection efficiency
was above \SI{99}{\percent} for single muon events, with a time resolution of \SI{350}{\pico\second}.

The NA48/2 main trigger, used in the presented results, was optimised for three-track vertex topologies. Coincidence of
hits in the two HOD planes was required in at least two of the 16 segments, or at least one segment accompanied by
at least two clusters of energy deposition in the LKr. Fast algorithms were run to reconstruct tracks from the DCH hits
and accept events with a three-track closest distance of approach below \SI{5}{\cm}. Alternatively, the track assumed to
be a \(\pi^\pm\), should have an energy \(E^* < \SI{230}{\mega\eV}\) in the \(\SI{60}{\giga\eVperc}~K^\pm\) rest frame.
This condition suppressed the \(K^\pm\to\pi^\pm\pi^0~(E^*=\SI{248}{\mega\eV})\) while keeping the
\(K^\pm\to\pi^\pm\pi^0\pi^0\). For \nark the trigger operating in minimum-bias mode was optimised for decays into
electrons: at least one coincidence of hits in the HOD, bounds on the hits multiplicity in the DCH, at least
\SI{10}{\giga\eV} of energy deposit in the LKr and at least one track with \(E/p > 0.6\) and
\(p\in\SIrange{5}{90}{\giga\eVperc}\) (with the energy \(E\) reconstructed in the LKr and the momentum \(p\)
reconstructed in the DCH).

\section{$\pi^0$ electromagnetic transition form factor slope at the \nark experiment}
\label{sec:pi0dalitz}
As the \(\pi^0\) is produced in four of the main \(K^\pm\) decays, with about \SI{2e10} collected kaon decays \nark is
an ideal experiment to study the neutral pion. The Dalitz decay \(\pizerod \to \gamma e^+ e^-\) \cite{Dalitz1951} has a
branching fraction of \(\mathcal{B} = \SI{1.174(35)}{\percent}\) \cite{PDG2014} and proceeds through the
\(\pi^0\gamma\gamma\) vertex with one off-shell photon. The commonly used kinematic variables defined in terms of the
particle four-momenta are:
\begin{equation*}
    x = \left( \frac{M_{e e}}{m_{\pi^0}} \right)^2
    = \frac{(p_{e^+} + p_{e^-})^2}{ m_{\pi^0}^2}, \qquad
    y = \frac{2 \, p_{\pi^0} \cdot \left( p_{e^+} - p_{e^-} \right)}{m_{\pi^0}^2
    (1-x)} \;,
\end{equation*}
where \(p_{\pi^0}, p_{e^+}, p_{e^-}\) are respectively the \(\pi^0\) and \(e^\pm\) four-momenta, \(m_{\pi^0}\) is
the mass of the \(\pi^0\), and \(M_{ee}\) is the \(e^+e^-\) invariant mass. The physical region is given by
\begin{equation*}
    r^2 = \left(\frac{2 m_e}{m_{\pi^0}}\right)^2 \leq x \leq 1, \quad |y| \leq
    \sqrt{1 - \frac{r^2}{x}} \;.
\end{equation*}
The \(\pizerod\) differential decay width normalised to the \(\pi^0_{2\gamma} \to \gamma\gamma\) decay width reads:
\begin{equation*}
    \frac{1}{\Gamma(\pi^0_{2\gamma})} \frac{\text{d}^2 \Gamma(\pi^0_D)}{\text{d}x \text{d}y} =
    \frac{\alpha}{4\pi} \frac{(1-x)^3}{x} \left(1 + y^2 + \frac{r^2}{x}\right) \; \left(1+\delta(x,y)\right) \;
    \left|\mathcal{F}(x)\right|^2 \;,
\end{equation*}
where \(\mathcal{F}(x)\) is the semi-off-shell electromagnetic transition form factor (TFF) of the \(\pi^0\) and
\(\delta(x,y)\) encodes the radiative corrections. 

The TFF is usually expanded as \(\mathcal{F}(x) = 1+ax\) where \(a\) is the form factor slope parameter. This
approximation is justified by the smallness of this parameter. It has been studied first in the vector meson dominance
(VMD) model \cite{GellMann1961} where it is dominated by the \(\rho\) and \(\omega\) mesons, resulting in a value
\(a\approx m^2_{\pi^0} \left(m_\rho^{-2}+m_\omega^{-2}\right)/2 \approx 0.03\). Further studies extending the VMD model
\cite{Lichard2011}, or using different frameworks \cite{Kampf2006,Husek2015,Masjuan2012,Hoferichter2014} are in
agreement with the original value. 

Another crucial aspect is the inclusion of the radiative corrections in the differential rate as they are of the same
size as the TFF. The total radiative corrections have been first studied in \cite{Joseph1960}, but the first
study of the corrections on the differential rate is done in \cite{Lautrup1971} in the soft-photon approximation. It has
been later extended in \cite{Mikaelian1972}. The most recent contribution \cite{Husek2015_rad}, triggered by this work,
includes new contributions and a code implemented at the MC event generator level for the generation of radiative
photon from the internal bremsstrahlung contribution.

\subsection{Event selection}
The selection is tuned to select a pure Dalitz decay sample from the dominant \(K^\pm \to \pi^+ \pi^0_D\) decay chain
(\(K_{2\pi D}\)), featuring two same-sign tracks, one opposite-sign track and a photon. Exactly one
three-track vertex must be reconstructed within the fiducial decay region and no additional track is allowed. The impact point of the three tracks
should be in the acceptance of the DCH chambers and separated by at least \SI{2}{\cm} in the first one. The momenta are
required to be in the range \SIrange{2}{74}{\giga\eVperc}. A single cluster of energy deposition with more than
\SI{2}{\giga\eV} of energy, and separated by at least \SI{20}{\cm} from the same-sign tracks and \SI{10}{\cm} from the
remaining track is allowed in the event. It is used to reconstruct a photon candidate assuming it originates from the
vertex. The total reconstructed momentum of the three tracks and the photon candidate should be in the range
\SIrange{70}{78}{\giga\eVperc} and the squared total transverse momentum with respect to the nominal beam axis \(p_t^2
< \SI{5e-4}{\left(\giga\eVperc\right)\squared}\). The mass assignment to the tracks is done using the kinematic
properties of this decay chain. The opposite-sign track is assigned an electron mass and two hypotheses are built for
the possible mass assignments of the same-sign tracks. Only events with at most one of the hypotheses satisfying the
following conditions on the reconstructed \(\pi^0\) and \(K^\pm\) masses and \(x,y\) kinematic variables are selected:
\(M_{ee\gamma}\in\SIrange{115}{145}{\mega\eVperc\squared} , \;
M_{\pi^\pm\pi^0}\in\SIrange{465}{510}{\mega\eVperc\squared}\), and \(x,|y|<1\). The trigger conditions described in
section \ref{sec:intro} must be reproduced, but with tighter criteria to eliminate edge effects due to different
resolution between online and offline analysis. The total LKr electromagnetic energy should be higher than
\SI{14}{\giga\eV} and at least one of the tracks is required to have \(p>\SI{5.5}{\giga\eVperc}\) and \(E/p>0.6\). Due
to the acceptance not being well reproduced in the simulation for events with  low \(x\), the signal region is defined
as \(x>0.01\), equivalent to \(M_{ee}>\SI{13.5}{\mega\eVperc\squared}\).

The final selected sample amounts to \num{1.1e6} events. The acceptances, evaluated from Monte-Carlo simulations, are
\SI{1.81}{\percent} for \(K_{2\pi D}\) and \SI{0.02}{\percent} for \(K^\pm \to \pi^0_D \mu^\pm \nu\) (\(K_{\mu 3 D}\)).

\subsection{Preliminary result}
A \(\chi^2\) fit of the \(x\) distribution of data and simulation is used to extract the TFF slope value. An
equipopulous binning is used. The different hypotheses are tested by reweighing the MC events simulated with
a known slope  \(a_\text{sim}=0.032\). The main systematic uncertainties arise from the simulation of the beam
spectrum and from the calibration of the spectrometer global momentum scale.

\begin{figure}[h]
	\centering
	\subfloat{ \includegraphics[width=0.49\columnwidth]{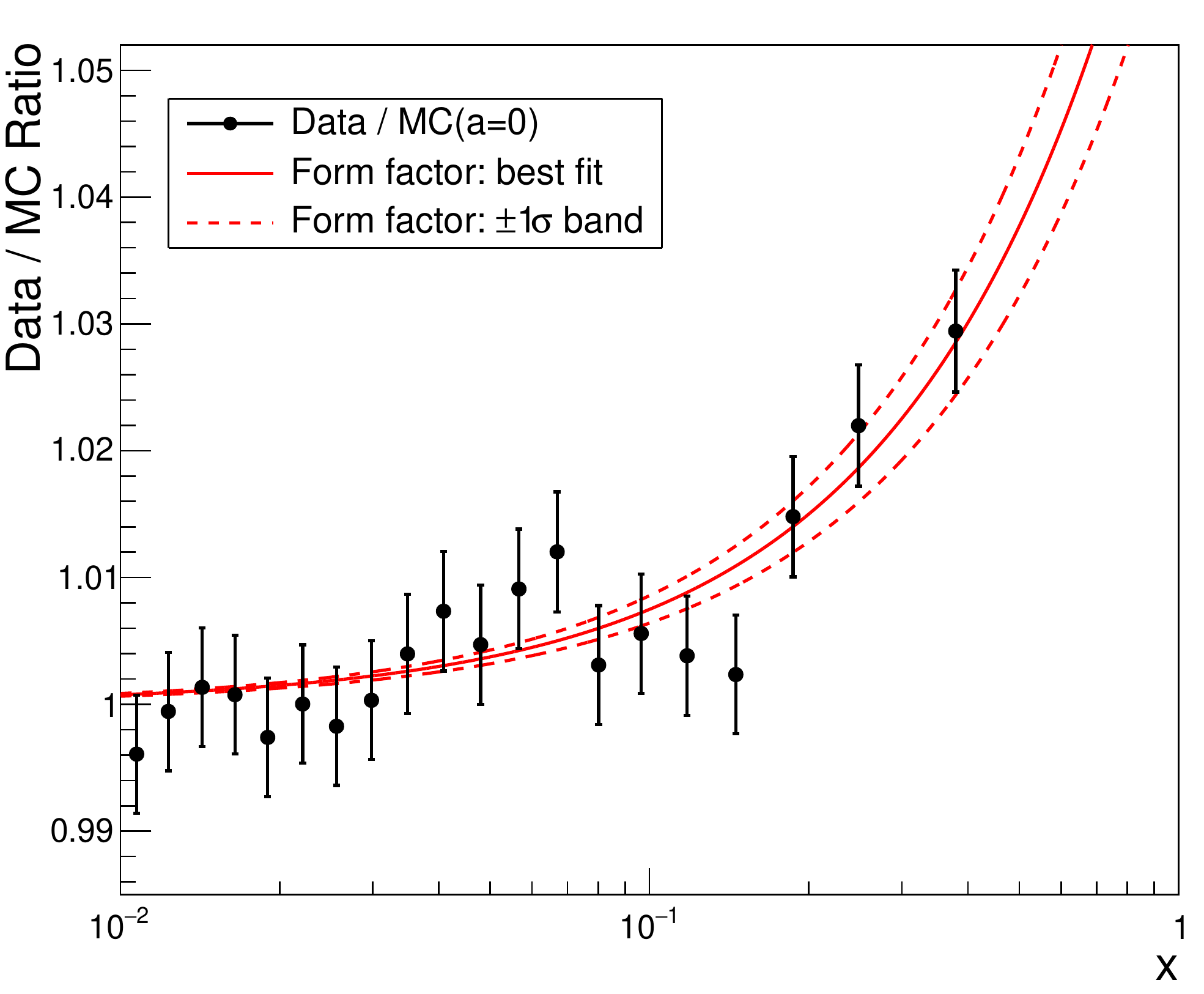} }
	\subfloat{ \includegraphics[width=0.48\columnwidth]{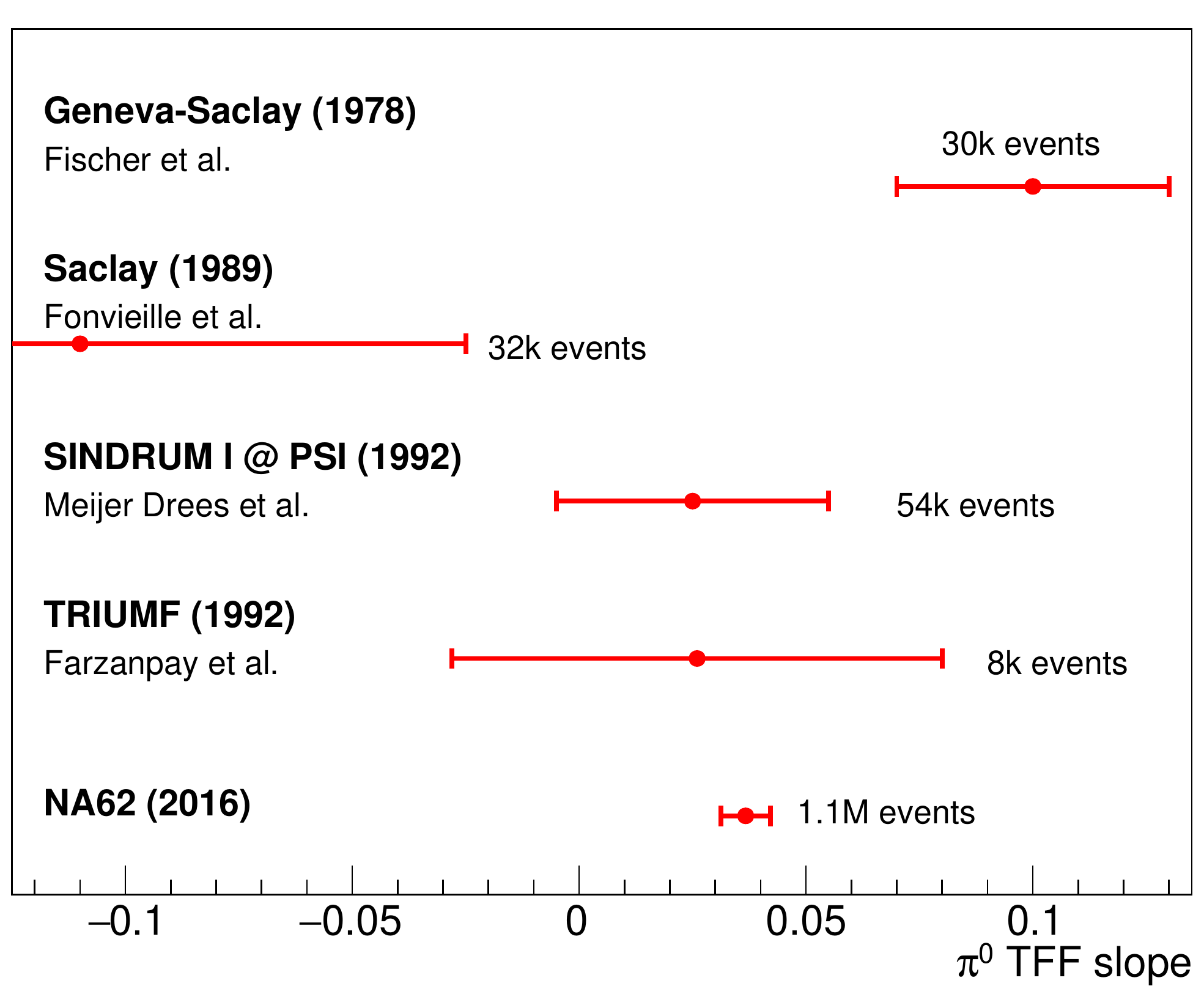} }
	\caption[Fit illustration and comparison with previous experiments.]
		{(a) Illustration of the fit result showing the ratio data/MC, with the MC sample weighted to obtain \(a=0\). The
		events are divided into 20 equipopulous bins with the points at the barycentre of each bin. The solid line
		represents the TFF function with a slope value equal to the fit central value. Dashed lines indicate the
		\(1\sigma\) band. (b) Comparison of this result with previous experiments measuring the TFF slope from \(\pi^0_D\)
		decays.}
	\label{fig:pi0d_result}
\end{figure}

The preliminary result obtained is:
\begin{equation*}
a = (3.70 \pm 0.53_\text{stat} \pm 0.36_\text{syst})\times 10^{-2} \;,
\end{equation*}
with \(\chi^2/\text{ndf} = 52.5/49\), which has a \textit{p}-value of \num{0.34}. This measurement represents a
significant measurement of a positive \(\pi^0\) electromagnetic TFF slope of more than \(5\sigma\) in the time-like
region of momentum transfer. An illustration of the best fit result and the comparison with previous measurements from
\(\pi^0\) Dalitz are shown in Fig. \ref{fig:pi0d_result}.

\section{Searches for lepton number violation and resonances in $K^\pm\to\pi\mu\mu$ at the NA48/2 experiment}
\label{sec:LFV}
Lepton number violation has never been observed. However heavy Majorana neutrinos are possible
mediators for such processes. The \(\nu\text{MSM}\) model \cite{Asaka2005} introduces three heavy sterile neutrinos \(N_i\).
The lightest one \(N_1\) has a mass of \(\mathcal{O}(\SI{}{\kilo\eVperc\squared})\) and is also a candidate for dark
matter. The two others have masses ranging from \SI{100}{\mega\eVperc\squared} to few \SI{}{\giga\eVperc\squared}. The
model can explain baryon asymmetry through CP-violating sterile neutrino oscillations and mixing with the active
neutrinos, and the low neutrino masses through see-saw mechanism. Effective vertices with the \(W^\pm,Z\) and the SM
leptons can be built, with a mixing matrix \(U\) describing the mixing between sterile and SM neutrinos. A simple
extension of this model with a real scalar field (inflaton \(\chi\)) \cite{Asaka2006} can also explain the homogeneity
and isotropy of the universe on large scales and the existence of structures on smaller scales.

The particular case of the lepton number violating (LNV) decay \(K^\pm\to\pi^\mp\mu^\pm\mu^\pm\) is investigated, with
the data sample of NA48/2 containing \(\sim \num{1.6e11}~K^\pm\) decays in vacuum. It can be mediated by the sterile
neutrinos if they are produced in the kaon decay as \(K^\pm\to\mu^\pm N\), then subsequently decay as
\(N\to\pi^\mp\mu^\pm\). For a neutrino produced as a resonance (indicated as \(N_4\)) with a mass \(m_4\), the possible
mass range is \(m_\pi+m_\mu < m_4 <m_K-m_\mu\). The total branching fraction for this channel would be:
\begin{equation*}
\mathcal{B}(K^\pm\to\pi^\mp\mu^\pm\mu^\pm) = \mathcal{B}(K^\pm\to\mu^\pm N)\times\mathcal{B}(N\to\pi^\mp\mu^\pm) \sim
\left|U_{\mu4}\right|^4\;,
\end{equation*} 
where \(U_{\mu4}\) is the mixing matrix element corresponding to the resonant sterile neutrino and the muon neutrino.
The current limit on this branching ratio is \(\mathcal{B} < \num{1.1e-9}\) at \npcl \cite{PDG2014}. The same neutrino
can also be seen in the lepton number conserving (LNC) channel \(K^\pm\to\pi^\pm\mu^\pm\mu^\mp\) if it decays as
\(N\to\pi^\pm\mu^\mp\). The inflaton in the mass range \(2m_\mu < m_\chi < m_K-m_\pi\) can appear in a similar way in
the decay \(K^\pm\to\pi^\pm \chi\) and subsequently decays into a muon pair. The production branching fraction is given
by \cite{Bezrukov2010}:
\begin{equation*}
\mathcal{B}(K^\pm\to\pi^\pm\chi) =
\num{1.3e-3}\left(\frac{\sqrt{\left(m_K^2-m_\pi^2-m_\chi^2\right)^2-4m_\pi^2m_\chi^2}}{m_K^2}\right)\theta^2\;,
\end{equation*}
where 
\(\theta\) is the inflaton mixing with the Higgs boson.

\subsection{Event selection}
The selection is developed based on the MC simulation of \(K^\pm\to\pi^\mp\mu^\pm\mu^\pm\),
\(K^\pm\to\pi^\pm\mu^\pm\mu^\mp\) and \(K^\pm\to\pi^\pm\pi^+\pi^-\) (\(K_{3\pi}\)) decays. This last channel is the main
background when \(\pi^\pm\) are mis-identified as \(\mu^\pm\). Since the event topology is similar, the systematic
effects due to trigger efficiencies cancel at first-order. The selection is based on a three-track vertex topology,
where the vertex is reconstructed inside the decay volume. The total reconstructed momentum of the three tracks must be
compatible with the nominal beam momentum and with no transverse momentum with respect to the nominal beam direction. Contrary to
the selection described in section \ref{sec:pi0dalitz}, the \(\pi,\mu,e\) separation is done using the measured \(E/p\)
ratio of the tracks, thus requiring the track impact points on the LKr plane to be in the acceptance. To obtain an
additional separation factor, the \(\mu^\pm\) are required to have associated hits in the MUV1 and MUV2 while the
\(\pi^\pm\) cannot be associated with any pair of hits from different MUV planes. 

For the LNV channel, the muons are required to have the same sign, with an opposite sign pion. A blind analysis is
performed and the tuning of the selection is done based on the control region defined as
\(M_{\pi\mu\mu}<\SI{480}{\mega\eVperc\squared}\), where \(M_{\pi\mu\mu}\) is the reconstructed invariant mass of the
pion and the two muons. The signal region is defined as \(|M_{\pi\mu\mu}-m_K|<\SI{5}{\mega\eVperc\squared}\).

Conversely, for the LNC channel the muons tracks are requested to have opposite sign. The signal region is defined
as \(|M_{\pi\mu\mu}-m_K|<\SI{8}{\mega\eVperc\squared}\). A total of 3489 candidate events are selected with a background
estimated at \(\SI{0.36(10)}{\percent}\).

\subsection{Upper limit on the lepton number violating decay $K^\pm\to\pi^\mp\mu^\pm\mu^\pm$}
After the final selection, a single event is observed in the signal region while the expected number of background
events is \(N_\text{exp} = 1.163\pm0.867_\text{stat}\pm0.021_\text{ext}\pm0.116_\text{syst}\). The obtained upper limit
at \npcl on the branching fraction is
\begin{equation*}
\mathcal{B}(K^\pm\to\pi^\mp\mu^\pm\mu^\pm) = \frac{N_\text{est}}{N_K A_\text{sig}}
\mathcal{B}(K_{3\pi})< \num{8.6e-11}\;, 
\end{equation*}
where \(N_\text{est}\) is the upper limit on the number of signal events, \(N_K\) is the total number of reconstructed
\(K^\pm\) decays in the fiducial decay region and \(A_\text{sig} = \SI{20.62(1)}{\percent}\) is the acceptance
for the signal estimated from the MC simulation. The upper limit at \npcl on \(N_\text{est}\) is computed using the
Rolke-Lopez statistical treatment \cite{Rolke2001, Rolke2005} from the number of observed events in the signal region,
the number of background events expected from the MC simulation and the associated uncertainty. This limit represents
an improvement by a factor of 10 with respect to the previous measurement \cite{PDG2014}.

\subsection{Search for two-body resonances}
A scan for a resonance of mass \(M_\text{res}\) in the reconstructed \(\pi^\pm\mu^\mp\) invariant mass \(M_{\pi\mu}\)
of both LNV and LNC channels, and in the \(\mu^\pm\mu^\mp\) invariant mass \(M_{\mu\mu}\) of the LNC channel is done.
The step and width of the search windows are determined by the mass resolution \(\sigma({M_\text{res}})\) at the tested
value. The mass step between two mass hypotheses is \(0.5\sigma(M_\text{res})\) and the window centered on
\(M_\text{res}\) has a width of \(2\sigma(M_\text{res})\). As the window is four times larger than the mass step, the
result for neighbouring mass hypotheses is highly correlated. The upper limit at \npcl on the product of branching
fractions is computed as
\begin{equation*}
\text{UL}(\mathcal{B}(K^\pm\to p_1 X)\mathcal{B}(X\to p_2 p_3)) = \frac{N_\text{est}}{N_K A_\text{sig}}\;,
\end{equation*}
where \((p_1,p_2,p_3) \to (\mu^\pm\pi^\mp\mu^\pm, \mu^\pm\pi^\pm\mu^\mp, \pi^\pm\mu^+\mu^-)\) and \(N_\text{est}\) is
the upper limit at \npcl on the number of signal events in the mass window. This is again estimated by the Rolke-Lopez
method from the number of observed events and the number of expected background events in each mass window. As a result
of the three-track vertex requirement, the acceptance depends also on the lifetime of the resonance \(\tau_\text{res}\)
and drops to zero as the distance travelled from the initial vertex by the resonant particle increases. 

\begin{figure}
	\centering
	\subfloat{ \includegraphics[width=0.48\columnwidth]{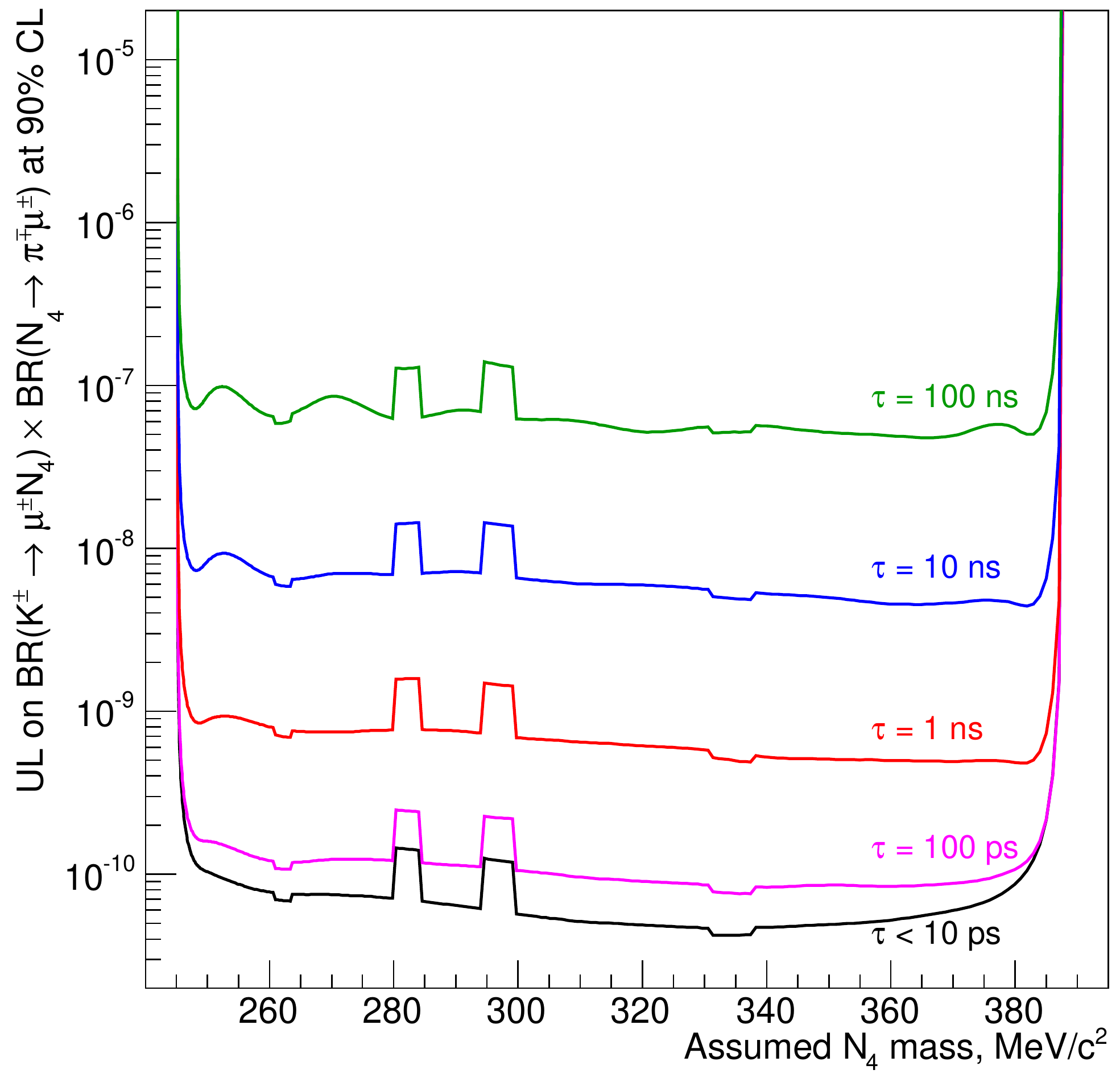} }
	\subfloat{ \includegraphics[width=0.48\columnwidth]{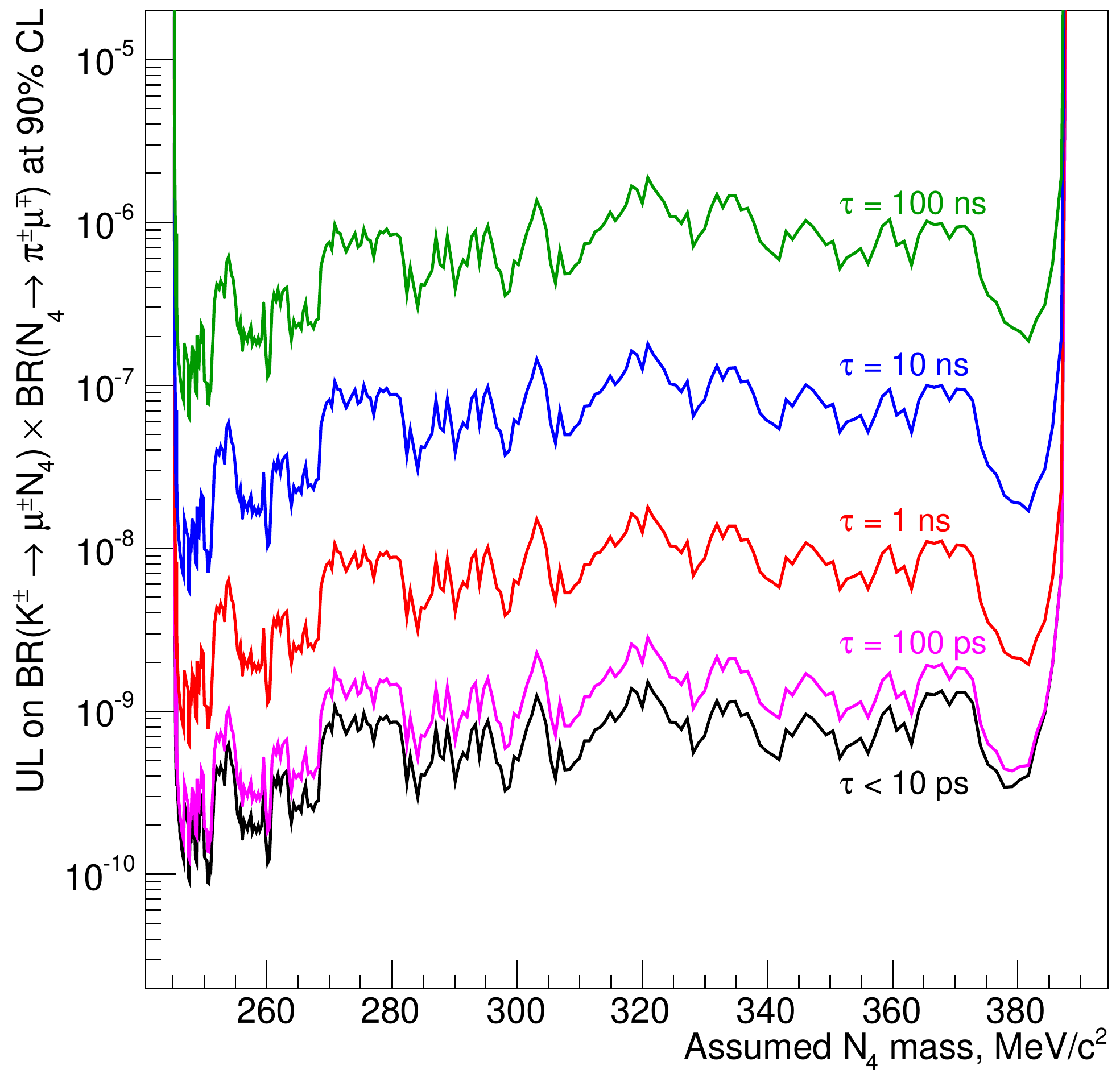} }\\
	\subfloat{ \includegraphics[width=0.48\columnwidth]{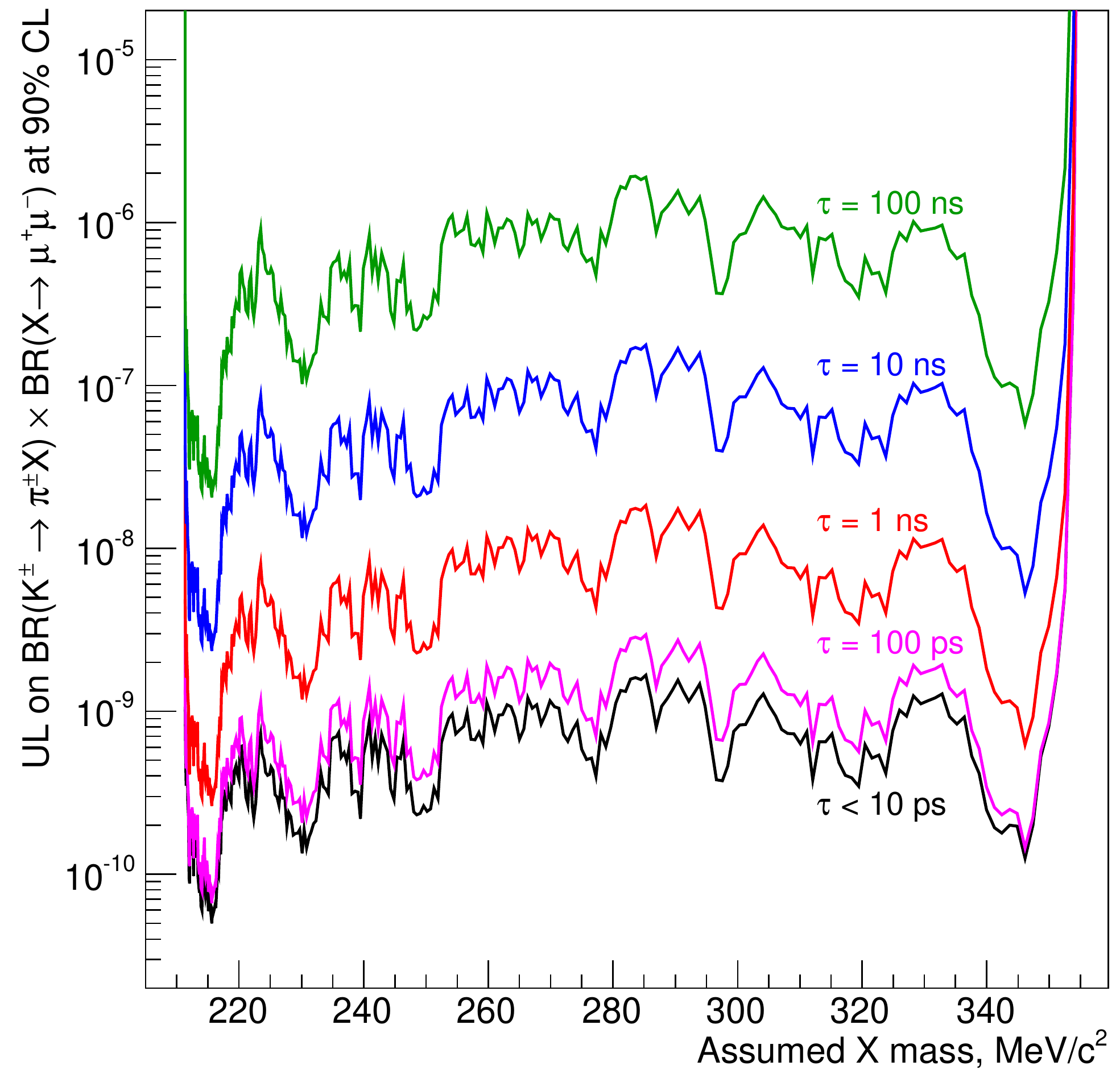} }
	\caption[Upper limits on branching fraction product]
		{Obtained upper limits at \npcl on the product of branching fractions as a function of the resonance mass for
		different lifetime of the resonant particle. (a) \(\mathcal{B}(K^\pm\to\mu^\pm
		N_4)\times\mathcal{B}(N_4\to\pi^\mp\mu^\pm)\) (b) \(\mathcal{B}(K^\pm\to\mu^\pm
		N_4)\times\mathcal{B}(N_4\to\pi^\pm\mu^\mp)\) (c) \(\mathcal{B}(K^\pm\to\pi^\pm
		N_4)\times\mathcal{B}(N_4\to\mu^\pm\mu^\mp)\).}
	\label{fig:lfv_result}
\end{figure}

A total of 284 mass hypotheses are tested on the LNV \(M_{\pi\mu}\) spectrum to search for \(K^\pm\to\mu^\pm
N_4;\; N_4\to\pi^\mp\mu^\pm\) decays. In this case two choices are possible to build \(M_{\pi\mu}\), selecting one
of the muons or the other. The possibility closest to \(M_\text{res}\) is chosen. No signal is observed and an upper
limit of \(\mathcal{O}(\num{e-10})\) is set on the product of the branching ratios for heavy Majorana neutrinos with
\(\tau_{N_4}~<~\SI{100}{\pico\second}\). The LNC \(M_{\pi\mu}\) spectrum is tested with 280 hypotheses and no
\(K^\pm\to\mu^\pm N_4;\; N_4\to\pi^\pm\mu^\mp\) signal is observed. The upper limit is \(\mathcal{O}(\num{e-9})\) for
sterile neutrinos with \(\tau_{N_4} < \SI{100}{\pico\second}\). The last scan done on the \(M_{\mu\mu}\) spectrum of
the LNC channel to search for \(K^\pm\to\pi^\pm \chi;\; \chi\to\mu^\pm\mu^\mp\) is done with 267 hypotheses and no
signal is observed. The upper limit is \(\mathcal{O}(\num{e-9})\) for an inflaton with \(\tau_{\chi} <
\SI{100}{\pico\second}\). The ULs as a function of the resonance mass are shown in Fig. \ref{fig:lfv_result} for
several lifetime values up to \SI{100}{\nano\second}.

\section{Dark photon searches at the NA48/2 experiment}
\label{sec:dp}
An extra \(U(1)\) symmetry is the simplest hidden sector model that can be introduced as an explanation of the
excess of positrons observed in cosmic rays \cite{Ackerman2012, Adriani2013, Accardo2014}, or the muon gyromagnetic
ratio (\(g-2\)) anomaly \cite{Pospelov2009}. It gives rise to the dark photon \(A'\), a new vector gauge boson that
interacts with the visible sector through kinetic mixing with QED \cite{Holdom1986} and couples to the quarks and
leptons in the same way, with a coupling constant \(\varepsilon\).

The same NA48/2 data sample is used to search for dark photons generated in \(\pi^0\) decays in association with a SM
photon. The expected branching ratio depending on \(\varepsilon\) and the dark photon mass \(m_{A'}\) is
\cite{Batell2009}:
\begin{equation}
\label{eq:gammaA}
\mathcal{B}(\pi^0\to\gamma A') = 2\varepsilon^2 \left(1-\frac{m^2_{A'}}{m^2_{\pi^0}}\right)^3
\mathcal{B}(\pi^0\to\gamma\gamma)\;.
\end{equation}
The sensitivity decreases as the dark photon mass approaches the \(\pi^0\) mass due to the kinematic suppression factor
and limits the mass range for this analysis. In such case, the dark photon is below threshold for all decays
into charged SM fermions and the only accessible one is \(A'\to e^+e^-\). The expected total decay width then reads
\cite{Batell2009_2}:
\begin{equation*}
\Gamma_{A'} \approx \Gamma(A'\to e^+ e^-) = \frac{1}{3}\alpha\varepsilon^2 m_{A'} \sqrt{1-\frac{4m_e^2}{m^2_{A'}}}
\left(1+\frac{2m^2_e}{m^2_{A'}}\right)\;.
\end{equation*}
If \(\varepsilon^2>10^{-7}\) and \(m_{A'}>\SI{10}{\mega\eVperc\squared}\) the mean free path does not exceed
\SI{10}{\cm} and can be neglected, and the dark photon is assumed to decay promptly. The searched decay chain is then
\(K^\pm\to\pi^\pm\pi^0\;,\pi^0\to\gamma A'\;, A'\to e^+ e^-\) and the final state is therefore similar to the Dalitz
decay of the \(\pi^0\), which represents an irreducible background. A second selection focused on the decay chain
starting with the \(K^\pm\to\pi^0\mu^\pm\nu\) (\(K_{\mu3}\)) decay is also built.

\subsection{Event selection}
The logic of the events selection is similar to the \(\pi^0_D\) one presented in section \ref{sec:pi0dalitz} with one
notable difference: the \(E/p\) based particle separation method is used and therefore all tracks should be in the LKr
acceptance and well separated on the LKr plane. The distance \(D_{ij}\) between the track impact
points on the LKr are requested to be \(D_{ee}>\SI{10}{\cm}\) for electron candidates and \(D_{e\pi/\mu}>\SI{25}{\cm}\)
for electron to pion/muon candidates. The reconstructed masses should also satisfy
\(\left|M_{ee\gamma}-m_{\pi^0}\right|<\SI{8}{\mega\eVperc\squared}\) and \(\left|M_{\pi^\pm\pi^0}-m_K\right|
<\SI{20}{\mega\eVperc\squared}\). For the \(K_{\mu3}\) the requirement on the total invariant mass is replaced by \(|M^2_\text{miss}|<\SI{0.01}{(\giga\eVperc\squared)^2}\), where \(M^2_\text{miss} = (P_K - P_\mu - P_{\pi^0})^2\) is
the squared missing mass, \(P_\mu\) and \(P_{\pi^0}\) are the reconstructed momenta of the \(\mu^\pm\) and \(\pi^0\),
and \(P_K\) is the nominal beam momentum.

The total selected sample amounts to \num{1.67e7} fully reconstructed \(\pi^0_D\) candidates. The overall acceptances
are \SI{3.82}{\percent} for \(K_{2\pi}\) and \SI{4.20}{\percent} for \(K_{\mu3}\).

\subsection{Search for dark photon resonance}
The same mass scan framework using the Rolke-Lopez method as in section \ref{sec:LFV} is used to search for narrow peaks
in the reconstructed \(e^+e^-\) invariant mass spectrum \(M_{ee}\). The steps and width of the search windows are
determined by the resolution \(\sigma(M_{ee})\), which has been estimated from a \(K_{2\pi D}\) MC sample and is approximately given
by \(\sigma(M_{ee}) \approx 0.011\times M_{ee}\). The mass step for the scan is set to \(0.5\sigma(M_{ee})\) rounded
to the nearest multiple of \SI{0.02}{\mega\eVperc\squared}. The window is centered on the searched dark photon mass
with a width equal to \(3\sigma(M_{ee})\). A total of 404 \(m_{A'}\) hypotheses are tested in the range
\(\SI{9}{\mega\eVperc\squared} \leq m_{A'} \leq \SI{120}{\mega\eVperc\squared}\).

The \npcl limits are obtained from the relation
\begin{equation*}
\mathcal{B}(\pi^0\to\gamma A') = \frac{N_{A'}}{N_K}\left[\mathcal{B}(K_{2\pi})A(K_{2\pi}) +
\mathcal{B}(K_{\mu3})A(K_{\mu3})\right]^{-1}\;,
\end{equation*}
where \(N_{A'}\) is the number of dark photon candidates, \(N_K\) is the total number of kaon decays in the fiducial
region, and \(\mathcal{B}(K_{2\pi}), A(K_{2\pi}), \mathcal{B}(K_{\mu3})\) and \(A(K_{\mu3})\) are the branching fraction
and acceptances of the \(K_{2\pi}\) and \(K_{\mu3}\) decays respectively. The largest uncertainty (\SI{3}{\percent})
comes from the \(\pi^0_D\) branching fraction entering in the computation of \(N_K\) and is neglected. The \npcl upper
limits on the \(\varepsilon^2\) mixing parameter are calculated from eq. \ref{eq:gammaA}.

\begin{figure}
	\centering
	\includegraphics[width=0.49\columnwidth]{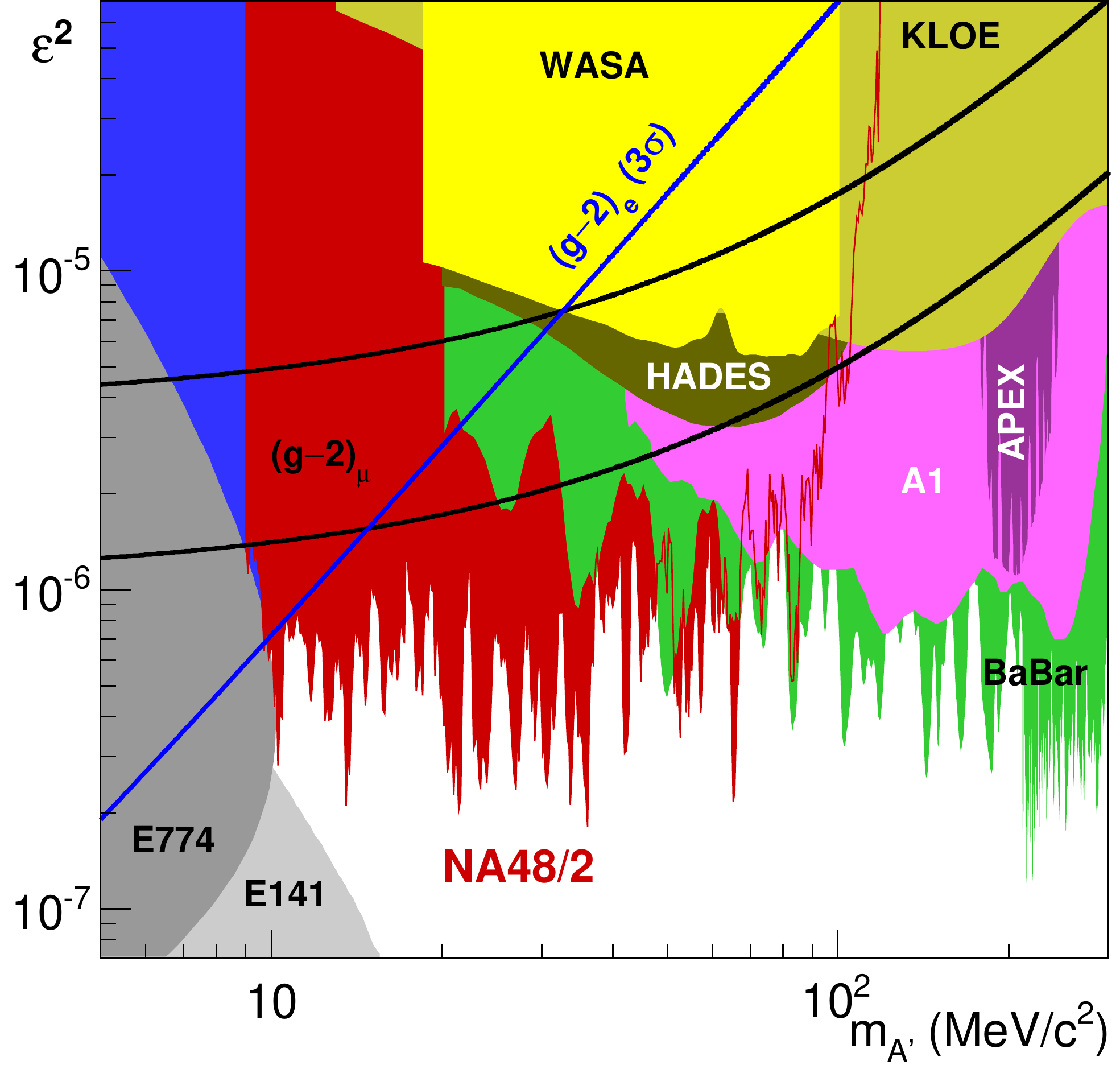}
	\caption[Upper limits]
		{Upper limits at \npcl on the mixing parameter \(\varepsilon^2\) for each dark photon mass hypothesis compared with
		exclusion limits set by previous experiments.}
	\label{fig:dp_result}
\end{figure}

No dark photon signal is observed in the scanned range. It constitutes an improvement on the existing limits in the
range \(m_{A'}\in \SIrange{9}{70}{\mega\eVperc\squared}\). The sensitivity is limited by the irreducible \(\pi^0_D\)
background and the upper limits are 2--3 orders of magnitude above the single event sensitivity. The resulting exclusion
limits associated to constraints from other experiments, as shown in Fig.~\ref{fig:dp_result}, completely rules out the
dark photon as an explanation for the muon (\(g-2\)) problem within the theoretical assumptions made in the analysis.

\section*{Conclusions}
The preliminary measurement of the electromagnetic transition form factor slope of the \(\pi^0\) using \num{1.05e6}
\(\pi^0\to e^+e^-\gamma\) decay collected at the NA62 experiment in 2007--08 is reported. The obtained result \(a =
(3.70 \pm 0.53_\text{stat} \pm 0.36_\text{syst})\times 10^{-2}\) is a significant measurement of more than \(5\sigma\)
of a positive slope in the time-like transfer momentum region.

Searches for the LNV \(K^\pm\to\pi^\mp\mu^\pm\mu^\pm\) decay and resonances in \(K^\pm\to\pi\mu\mu\) decays using the
2003--04 data sample collected at NA48/2 are also presented. No signals are observed. An upper limit on the branching
ratio of the LNV channel is set at \num{8.6e-11}, improving the previous best limit by an order of magnitude. Upper
limits are set on the products of branching ratios \(\mathcal{B}(K^\pm\to\mu^\pm
N_4)\mathcal{B}(N_4\to\pi^\mp\mu^\pm)\), \(\mathcal{B}(K^\pm\to\mu^\pm
N_4)\mathcal{B}(N_4\to\pi^\pm\mu^\mp)\) and \(\mathcal{B}(K^\pm\to\pi^\pm\chi)\mathcal{B}(\chi\to\mu^+\mu^-)\) as
functions of the resonance mass and lifetime. These limits are in the range \numrange{e-9}{e-10} for resonance lifetimes
below \SI{100}{\pico\second}.

Using the same data sample, upper limits are set on the mixing parameter \(\varepsilon^2\) of a dark photon \(A'\). The
dark photon is assumed to be produced in the decay \(\pi^0\to\gamma A'\) and decays only into an \(e^+e^-\) pair. The
limits are of the order of \num{e-6} in the mass range \(m_{A'}\in\SIrange{9}{70}{\mega\eVperc\squared}\), excluding
this dark photon as an explanation for the muon \((g-2)\) anomaly.

\end{document}